\documentclass[twocolumn,trackchanges]{aastex63}

\newcommand{\Msun}{\,\ensuremath{\mathrm{M}_\odot}}

\submitjournal{ApJL}

\shorttitle{Extended stellar halo formation in a UFD following a galaxy merger}
\shortauthors{Tarumi et al.}
\graphicspath{{./}{figures/}}

\begin{document}

\title{Formation of an extended stellar halo around an
ultra-faint dwarf galaxy \\following one of the earliest mergers from galactic building blocks}

\correspondingauthor{Yuta Tarumi}
\email{yuta.tarumi@phys.s.u-tokyo.ac.jp}

\author{Yuta Tarumi}
\affiliation{Department of Physics, School of Science, The University of Tokyo, Bunkyo, Tokyo 113-0033, Japan}

\author[0000-0001-7925-238X]{Naoki Yoshida}
\affiliation{Department of Physics, School of Science, The University of Tokyo, Bunkyo, Tokyo 113-0033, Japan}
\affiliation{Kavli Institute for the Physics and Mathematics of the Universe (WPI), UTIAS, The University of Tokyo, Chiba, 277-8583, Japan}
\affiliation{Research Center for the Early Universe, School of Science, The University of Tokyo, Tokyo, 113-0033, Japan}

\author[0000-0002-2139-7145]{Anna Frebel}
\affiliation{Department of Physics and Kavli Institute for Astrophysics and Space Research, Massachusetts Institute of Technology, Cambridge, \\
MA 02139, USA}



\begin{abstract}

Ultra-faint dwarf galaxies (UFDs) are promising observable proxies to building blocks of galaxies formed in the early Universe. We study the formation and evolution of UFDs using cosmological hydrodynamic simulations. In particular, we show that a major merger of two building block galaxies with $3,900\ \Msun\ \mathrm{and}\ 7,500 \Msun$ 
at the cosmic age of $510\ \mathrm{Myr}$ 
results in a system with an extended stellar distribution consistent with the de Vaucouleurs profile. 
The simulated galaxy has an average stellar metallicity of $\mbox{[Fe/H]}=-2.7$ and features 
a metallicity gradient. These results closely resemble the properties of a recently discovered UFD, Tucana~II, which is extremely metal-poor and has a spatially extended stellar halo 
with the more distant stars being more metal-poor.
Our simulation suggests that
the extended stellar halo of Tucana~II may have been formed through a past major merger. 
Future observational searches for spatially extended structures around other UFDs,
combined with further theoretical studies, will provide tangible measures of 
the evolutionary history of the ancient, surviving satellite galaxies. 

\end{abstract}





\section{Introduction} \label{sec:intro}

Ultra-faint dwarf galaxies (UFDs) are small satellite galaxies in the Local Group. They are very faint ($\leq 10^{5}\ \mathrm{L}_{\odot}$) and dark-matter dominated with mass-to-light ratios typically greater than $100$ \citep{Simon19_UFDreview}. 
UFDs host old stellar populations, and thus
it is thought that cosmic reionization quenched star formation in UFD progenitors by heating and evaporating the gas within their dark halos \citep{2014Brown_UFDquench}. 
Therefore, they are ideal fossils from the Universe before reionization.

Recently, \citet{2021Chiti_TucIIhalo} discovered an unusually extended stellar population around a UFD, Tucana~II (Tuc~II). Seven giant stars are located at distances larger than twice the half-light radius. The farthest star is more than 1\,kpc away from the center. Considering the small sample size of  19 giants, the overall size of the stellar density distribution is indeed significant. Importantly, the extended stars are likely members of Tuc~II because they have physical properties in agreement with those of the more centrally located stars: the systemic radial velocity ($-129.1\ \mathrm{km\ s^{-1}}$) and a very low metallicity ($[\mathrm{Fe}/\mathrm{H}] = -2.77$).  

Formation of such an extended stellar structure may have multiple origins. One possible mechanism is tidal disruption by the Milky Way (MW). A UFD  orbiting in the MW halo is disrupted by tidal forces of the MW's disk and dark matter halo the closer it is located, most notably for systems with $\lesssim 20\ \mathrm{kpc}$ (\citealt{2008Penarrubia}). But the distance of Tuc~II is 58\,kpc which is too far away to be significantly disrupted, and the overall stellar structure of Tuc~II appears unfavorable to the tidal disruption scenario \citep{2021Chiti_TucIIhalo}. Hence, another mechanism is needed to explain the particularly extended stellar structure of the galaxy.

An alternative mechanism is an early galaxy-galaxy merger. Mergers can disrupt the structure of either or both galaxies and 
cause the member stars to get scattered to large radii. In this {\it Letter}, we study this possibility in the context of early galaxy formation in the standard cosmological model. We follow the formation of UFD progenitors and investigate whether a merger can significantly change the structure of an early galaxy to produce the extended structure as observed for Tuc~II.

\section{Simulations of early galactic systems}

\begin{figure*}
    \centering
    \includegraphics[width=\columnwidth]{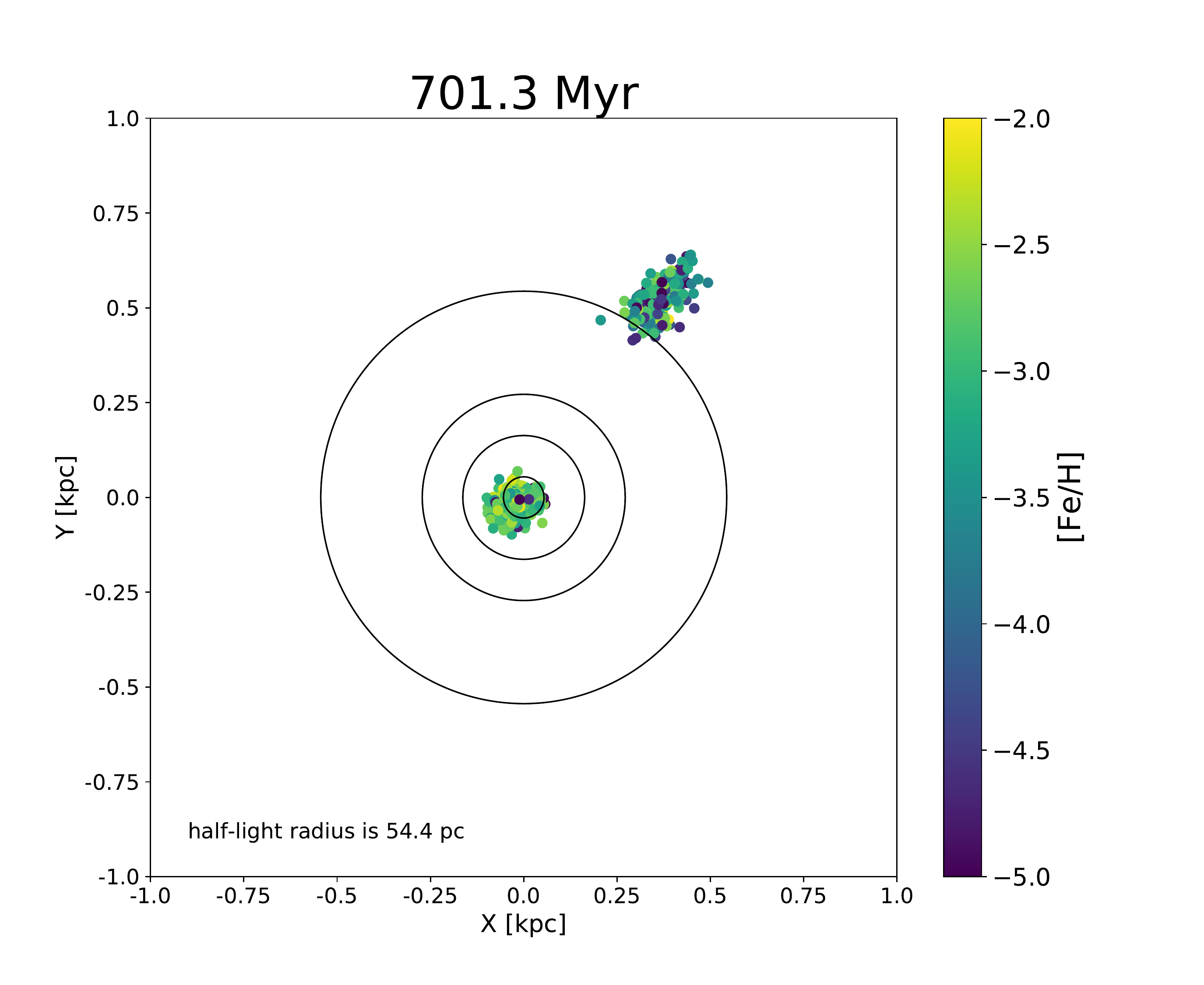}
    \includegraphics[width=\columnwidth]{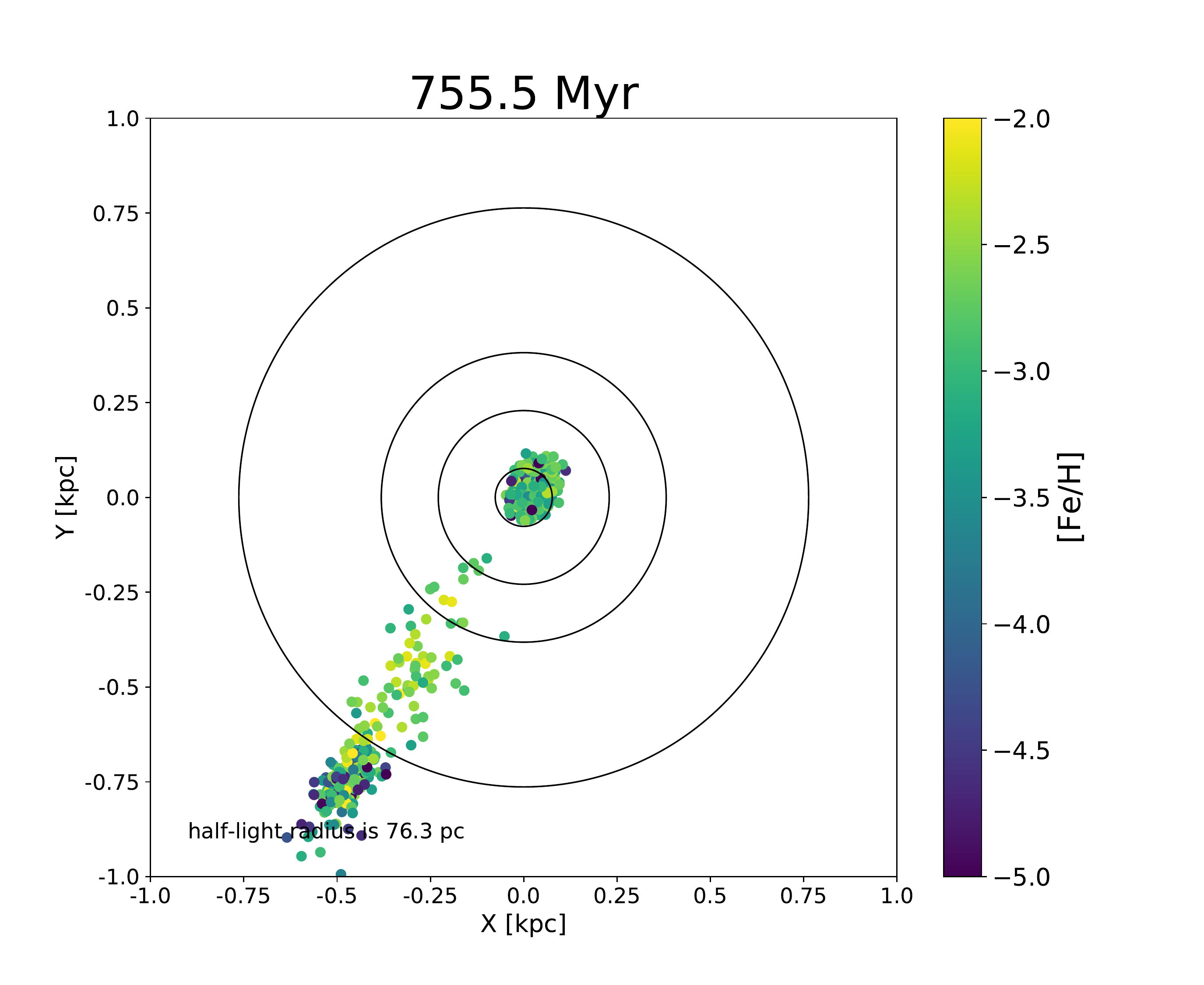}
    \includegraphics[width=\columnwidth]{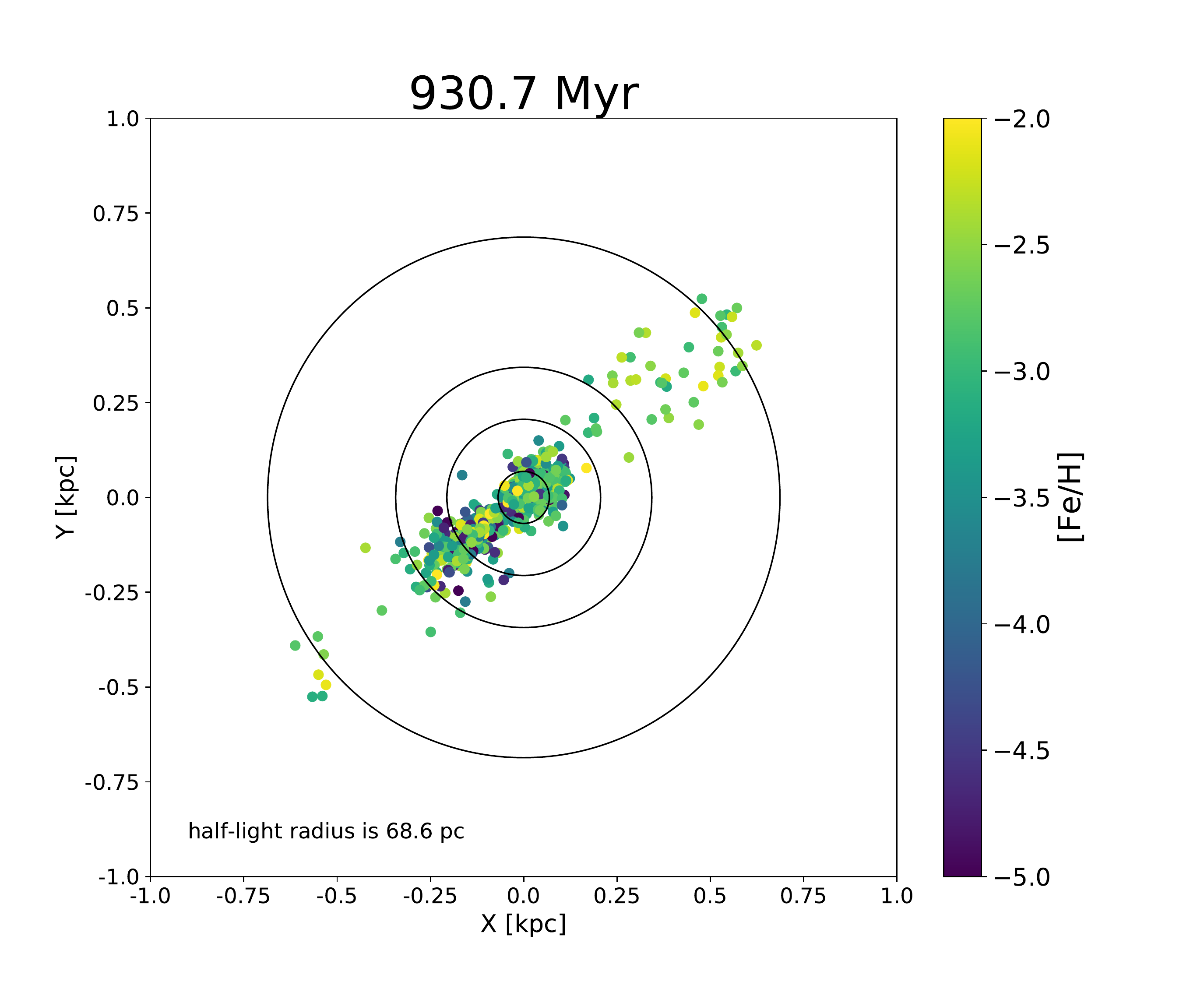}
    \includegraphics[width=\columnwidth]{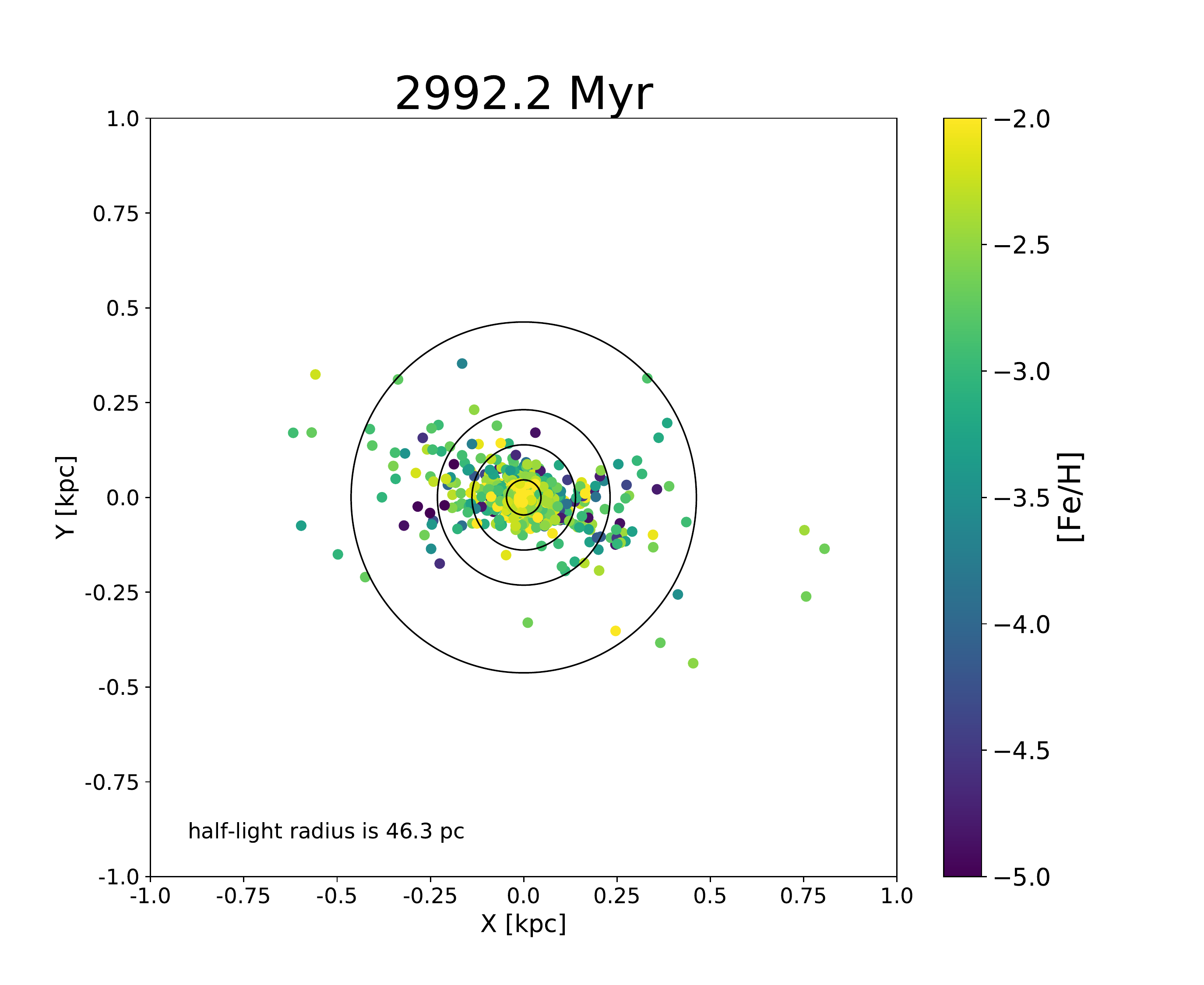}
    \caption{Spatial distribution of stars in our main simulation of galaxy merger. Time proceeds from top left to bottom right. We show a region of $2\ \mathrm{kpc}$ on a side. The four circles denote one, three, five, and ten half-light radii of the galaxy. The half-light radius is calculated assuming that the galaxy is observed in the direction of projection. 
    Top left: Stellar distribution before the close encounter of the two galaxies. 
    Top right: Stellar distribution after the disruption of the infalling galaxy. Bottom left: Stellar distribution $\sim 200\,\mathrm{Myr}$ after the merger. A bi-polar, bar-like structure develops in the direction of the merger. Bottom right: Stellar distribution $\sim 2,500\,\mathrm{Myr}$ after the merger. The extended bar-like structure still survives.}
    \label{fig:Spatial distribution}
\end{figure*}

\subsection{Galaxy formation model}

We use the parallel $N$-body/hydrodynamics code \textsc{arepo} \citep{Springel10_AREPO, 2020AREPOrelease}. The initial conditions are generated with MUlti-Scale Initial Condition generator \textsc{music} \citep{Hahn11_MUSIC}. We adopt the Planck 2018 cosmological parameters \citep{2020_Planck}: $\Omega_{m} = 0.315, \Omega_{b} = 0.049, \sigma_{8} = 0.810, n_{s} = 0.965, H_{0} = 67.4\ \mathrm{km\ s^{-1}\ Mpc^{-1}}$. 

We use a zoom-in technique to represent the UFD progenitors with sufficient resolution. Note that we aim to simulate the early formation and evolution of these galaxies. Therefore, we do not simulate the merger of these small galaxies with the MW halo. We first run two  cosmological parent simulations. The side lengths of the boxes are $1$ and $2$ $\mathrm{comoving}\ \mathrm{Mpc}\ h^{-1}$. We select five galaxies with dark matter halo masses of $\sim 10^{8} \Msun$ at redshift 8, which are good candidates for UFD progenitors \citep{Safarzadeh18_UFDselection}. 
We set a higher maximum refinement level in the simulation with the $2$ $\mathrm{comoving}\ \mathrm{Mpc}\ h^{-1}$ boxsize, so that the mass resolution is the same 
in all our zoom simulations. The mass of each dark matter particle is $102\ \Msun$ and the softening length is $10\ \mathrm{comoving\ pc}\ h^{-1}$. 
The typical mass of a gas cell is $19\ \Msun$ initially. The softening length of a gas cell with volume $V_\mathrm{gas}$ is $2.8\times V_\mathrm{gas}^{1/3}$, which is $25$ pc for a cell with the number density of $1\ \mathrm{cm^{-3}}$.

We do not adopt a multi-phase model for the interstellar medium (ISM) \citep{Springel03_WindParticles}
but implement a high density threshold model for star formation. 
We allow star formation to occur in gas cells with number densities of $n_\mathrm{gas} > 100\ \mathrm{cm^{-3}}$. 
The star formation rate (SFR) of each gas cell is $\mathrm(SFR) = 0.079\ m_\mathrm{g}/t_\mathrm{SF}$, where $m_\mathrm{g}$ is the mass of the gas cell and $t_\mathrm{SF} = (G\rho_\mathrm{g})^{(-1/2)}$ is the timescale of star formation. We only allow star formation in gas cells colder than $10,000\ \mathrm{K}$.
Stellar feedback is modeled in the same manner as in \citet{2020Tarumi_UFD_sprocess}.

\subsection{Final merged galaxy}
Among our five sample galaxies, only one experiences a merger after the onset of major star formation. The dark matter halo mass ($M_\mathrm{DM}$) and stellar mass ($M_\mathrm{*}$) of the two galaxies before the merger (at $510\ \mathrm{Myr}$) are $(M_\mathrm{DM}, M_\mathrm{*}) = (8.5\times 10^{7} \Msun, 3,900\ \Msun)$ and $(5.0\times 10^{7} \Msun, 7,500 \Msun)$, respectively. The halo mass ratio is $\sim 1.4$ which is typical for a major merger.
The medians and quartile deviations of [Fe/H] of the stars in the infalling are $-$3.06 and 1.36, and $-$2.72 and 0.70 for the central galaxy. 

The dark matter halo merger then occurs at the cosmic age of $510\ \mathrm{Myr}$. 
The close encounter of the stellar component occurs slightly after, at $700\ \mathrm{Myr}$. During and after the merger, $\sim 3,000 \Msun$ of stars are formed. They have higher median [Fe/H] and quartile deviation values of $-$2.29 and 0.56. At redshift 4.5 (corresponding to 1,370\,Myr),
the merged galaxy has achieved $(M_\mathrm{DM}, M_\mathrm{*}) = (2.5\times 10^{8},\Msun, 14,000\,\Msun)$. The star particles in the final merged galaxy have median and quartile deviation values of $\mbox{[Fe/H]} =-2.7$ and $0.82$. This agrees with the average metallicity found in UFDs, in particular with that if Tuc~II of  $\mbox{[Fe/H]} =-2.77$. 
By that time, an extended bar-like structure develops in the direction of the merger which spans over $\sim 1\,\mathrm{kpc}$. It survives for at least $2.5\,\mathrm{Gyr}$ after the merger. Interestingly, the two farthest confirmed members of Tuc~II are both located perpendicular to the direction of tidal disruption. The observed distribution may indicate that the stellar component of the galaxy also has a bar-like shape extended in this direction.

At this redshift we also switch off the hydrodynamic calculation and follow the evolution purely gravitationally in order to examine the stability and survival of the stellar system. We note that by the time of the switch-off, significant gas evaporation has already occurred, and thus, further star formation in the galaxy is unlikely. We then run the simulation for an additional 1.5\,Gyr until a cosmic age of 3\,Gyr.

\section{Properties of our simulated early, merged galaxy}

\begin{figure}
    \centering
    \includegraphics[width=\columnwidth]{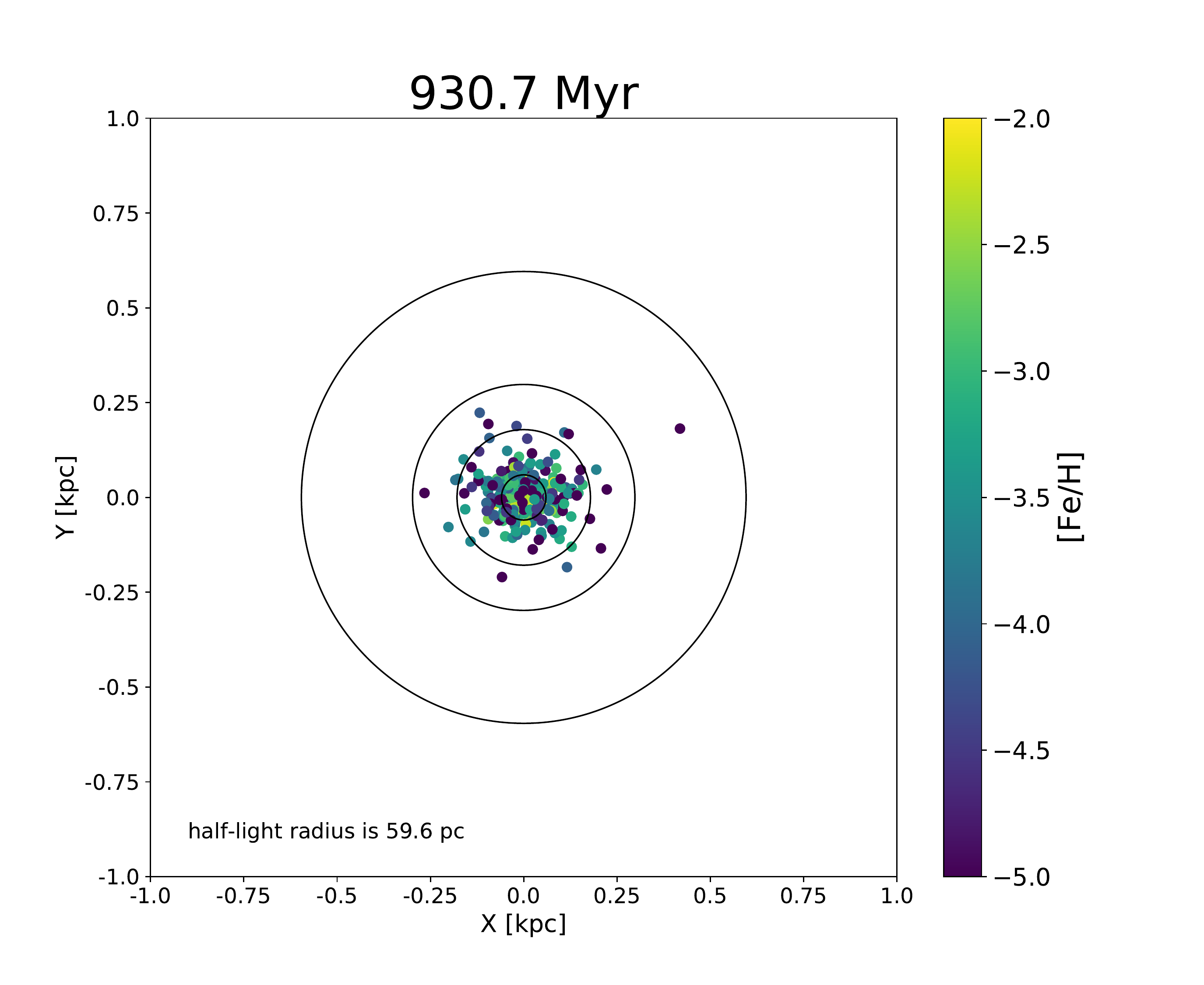}
    \caption{Stellar distribution of a typical non-merger galaxy. Symbols are the same as in Figure~\ref{fig:Spatial distribution}. Most stars are confined to within $5\ r_\mathrm{half}$ except for one star. The fraction of stars at $> 5\ r_\mathrm{half}$ is 0.3\% (see Figure~\ref{fig:Faraway fraction}).}
    \label{fig:no-merge galaxy}
\end{figure}

\subsection{Stellar distribution}

Figure~\ref{fig:Spatial distribution} presents the stellar distribution during the merger.  
We also show concentric circles with radii of 1, 3, 5, and 10 times the half-light radius $r_\mathrm{half}$, which corresponds to the median of the distances of star particles in the plane of the projection. 
Stars are colored according to their metallicities (see color bar on the right). 

Before the merger (top left panel), the two galaxies are very compact. They do not host stars at distances larger than $3\ r_\mathrm{half}$, and
the distributions are roughly spherical.
After the first close encounter (top right panel), the smaller galaxy is disrupted, and its member stars are then located
at various distances from the center of the main galaxy. 
At $\sim 200\,\mathrm{Myr}$ after the merger (bottom left panel), an extended, bar-like structure develops. The main constituents of the bar are the stars from the disrupted galaxy as well as those formed during the merger. The maximum extension of the bar-like structure is nearly $1\,\mathrm{kpc}$, with small variation 
from snapshot to snapshot.
The prominent bar-like structure is still present at $\sim 2,500\,\mathrm{Myr}$ after the merger (bottom right panel). The upscattered stars orbit in the halo. The farthest star is still within the virial radius of the galaxy implying that all stars are gravitationally bound. 
Beyond that, we find that ``dissipative" processes such as dynamical friction and redistribution of orbital energies do not  effectively occur
in the galaxy over $\sim 2$\,Gyr after the merger.

In Figure~\ref{fig:no-merge galaxy} we show the stellar distribution in another sample galaxy as a reference. This galaxy does not experience any merger after its major star formation episode. The stellar distribution appears spherical and is very compact. About $95$ percent of the stars are confined to $< 3\ r_\mathrm{half}$, and only one star ($0.3$ percent) is located at the distance larger than $5\ r_\mathrm{half}$. These features are consistent with an exponential surface density profile. This type of compact stellar structure is commonly found in our sample galaxies that do not experience mergers.

\subsection{Outer halo stars}
\begin{figure}
    \centering
    \includegraphics[width=\columnwidth]{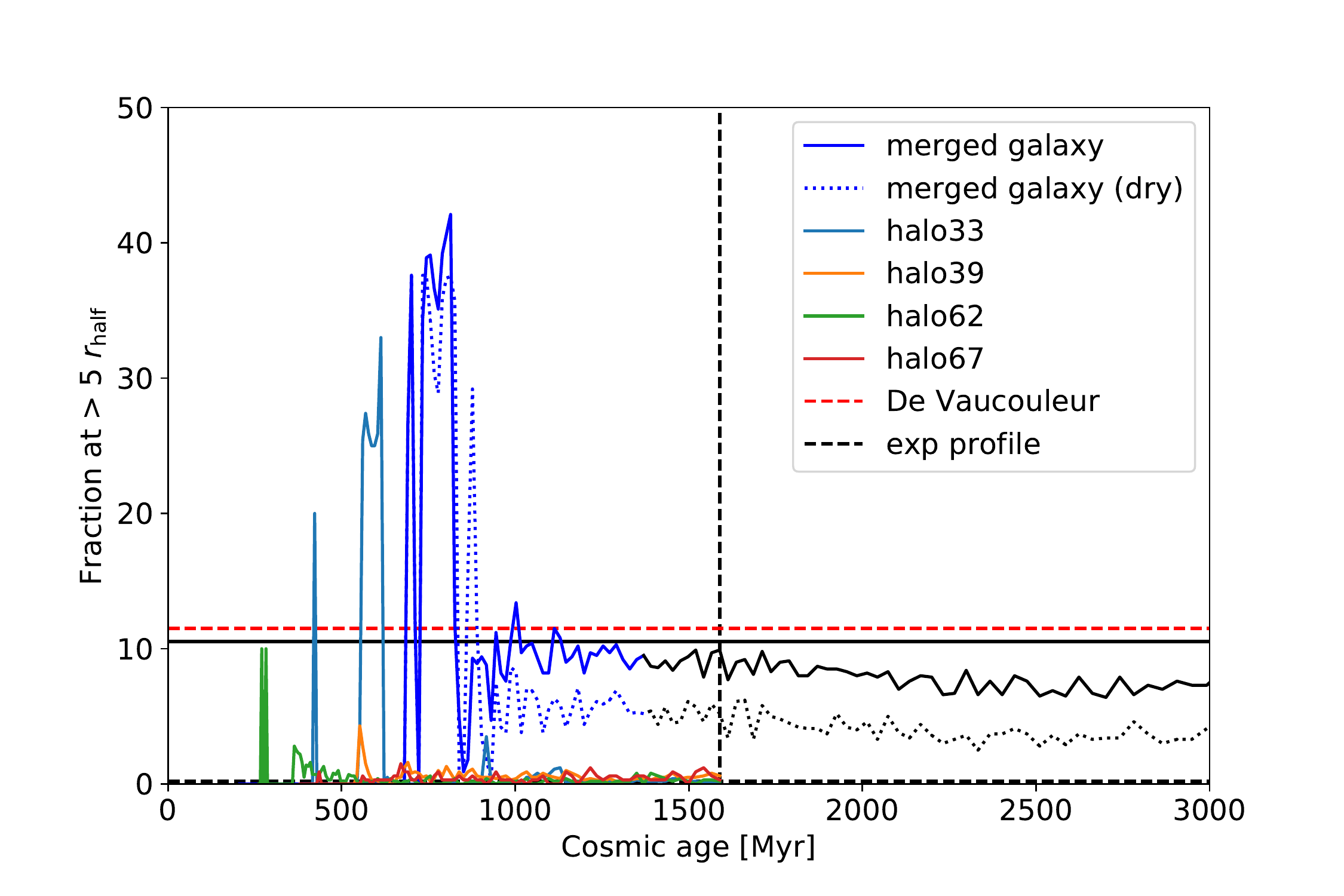}
    \caption{
    Fraction of spatially extended outer halo stars located at $> 5\ r_\mathrm{half}$
    in each galaxy ($f_5$). The blue solid line depicts the behavior of the merger galaxy within its first 1.4\,Gyr. 
    After $t=1.4$\,Gyr, we follow only the gravitational interaction of the system (see text for discussion). Then $f_5$ is depicted in black. Dotted lines depict results obtained when only stars formed prior to the merger are considered. Other galaxies from our simulation sample are shown with different colors. 
    The fraction suddenly rises when the two progenitor galaxies merge. After the merger, the outer halo fraction decreases and settles at $7-10$\%. 
    The Tuc~II fraction is $2/19=10.5$\% (black horizontal line) and similar to that of the de Vaucouleur profile ($f_5$ = 11.5\%; red dashed line). The fraction for the other, non-merger galaxies are small and converge to $f_5 \sim 0$, which is consistent with an exponential profile.} 
    \label{fig:Faraway fraction}
\end{figure}

To make a quantitative comparison of the stellar distributions out to large radii (``outer halo stars") between the merged galaxy and other galaxies in our simulation, we plot the fraction of stars at $> 5\ r_\mathrm{half}$ as a function of cosmic time in Figure~\ref{fig:Faraway fraction}. 
For non-merging galaxies, the fraction converges to $f_5 \sim 0$ after star formation ceases. This is consistent with exponential surface density profiles (0.2\%) that is also added in the Figure. This contrasts the observed fraction for Tuc~II of $f_5 = 2/19=10.5$\% which happens to be similar to a de Vaucouleurs profile (11.5\%).

The merged galaxy behaves differently. It experiences a sudden increase of its stellar fraction of spatially extended, outer halo stars during and after the merger. Eventually, that fraction decreases when the galaxy dynamically relaxes and $f_5$ settles at $\sim 10$ percent, around $\sim 500$\,Myr after the merger. Although the fraction decreases slightly with time, the difference between the merged galaxy and the other galaxies remains large even after 2.5\,Gyr of evolution. The fraction of outer halo stars in the merged galaxy is consistent with the observed fraction found for Tuc~II.

\subsection{Metallicity gradient}
\begin{figure}
    \centering
    \includegraphics[width=\columnwidth]{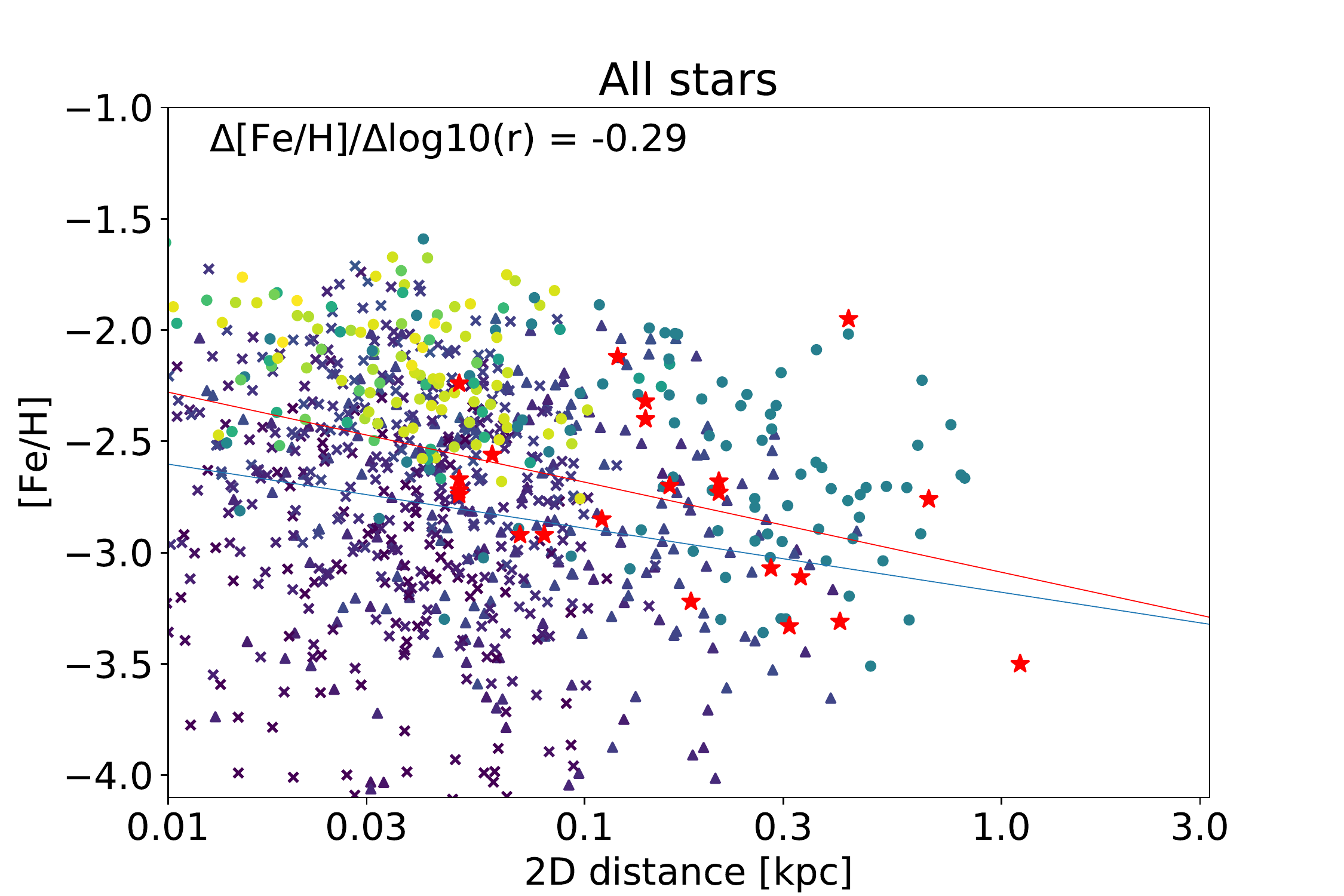}
    \includegraphics[width=\columnwidth]{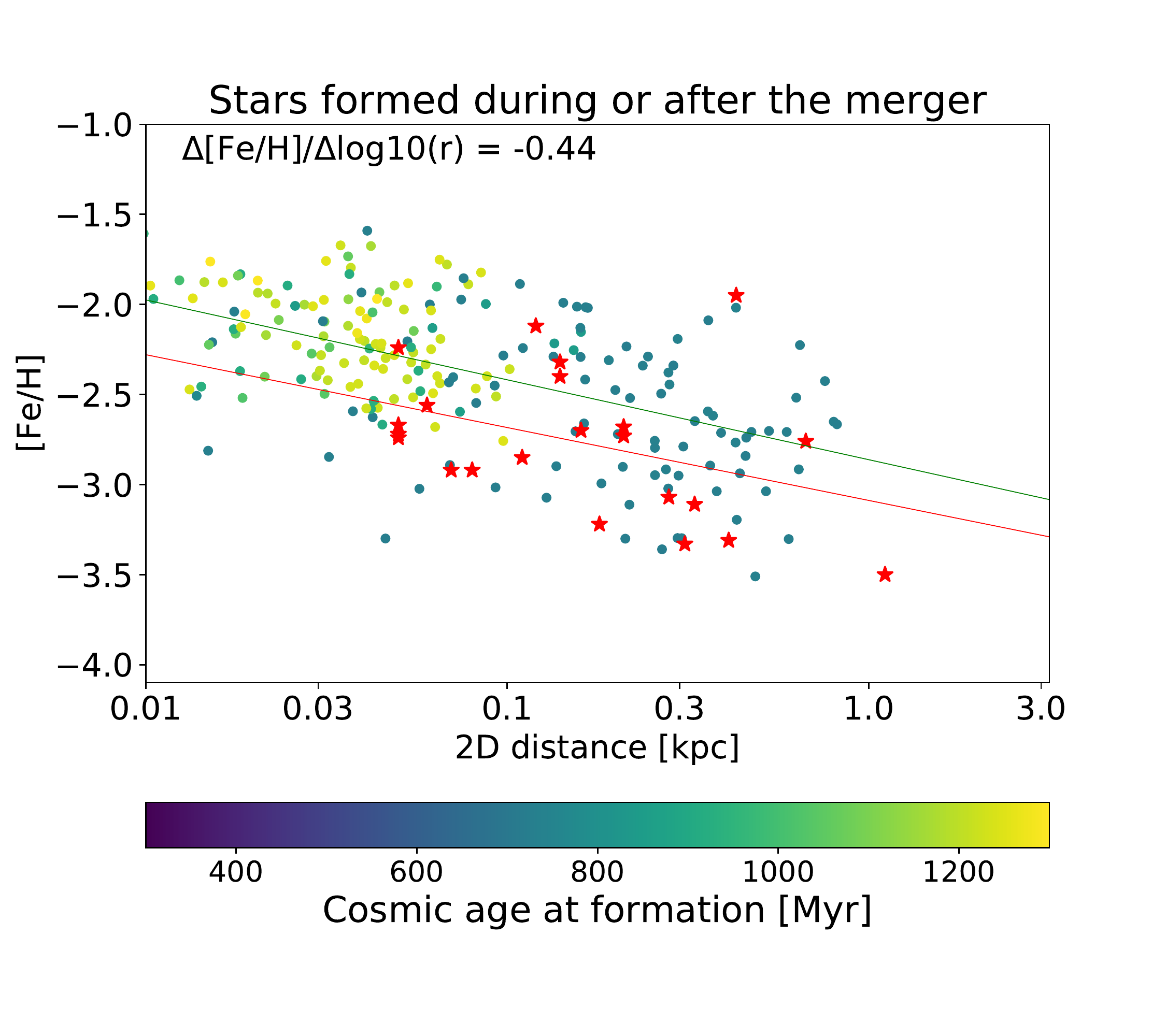}
    \caption{Stellar metallicity distribution of our merger galaxy sample. Top panel: Stars in the merger galaxy. Red symbols indicate the observed stars in Tuc~II, and the other symbols (triangles, crosses, and circles) depict star particles in the simulation. Symbol colors represent their formation times. Stars in the accreting galaxy are shown as triangles, those in the central galaxy as crosses, and those formed during or after the merger as circles. The blue line shows the result of the linear regression of the simulated star particles. Stars with [Fe/H] $< -4.0$ are excluded from the regression analysis. Bottom panel: Stars formed during or after the merger (i.e., green dots in the top panel). Here the color depicts stellar formation time. The green line shows the result of the linear regression of this sample.}
    \label{fig:FeH_distance}
\end{figure}

Figure~\ref{fig:FeH_distance} shows simulated and observed stellar iron abundances [Fe/H] as a function of the projected distance from the center. 
In the top panel, we plot all stars whereas only stars formed during or after the merger are shown in the bottom panel. 
From linear regression we find a weak radial metallicity gradient to be present in the merged galaxy (solid lines). 
Note that the negative gradient persists in all the later snapshots not just the one shown in the Figure, although the slope fluctuates from about $-0.2\ \mathrm{dex\ kpc^{-1}}$ to $-0.8\ \mathrm{dex\ kpc^{-1}}$. 

In the bottom panel of Figure~\ref{fig:FeH_distance}, the color of each dot indicates the formation time of a star particle. Interestingly, the distribution of stars formed during or after the merger presents a even more pronounced metallicity gradient that is likely produced by a combination of the following two processes.
Most stars in the outskirt are formed shortly after the merger, while the metal-rich stars are formed ``in situ" at the center 
a few hundred Myr after the merger. 
The former population is scattered out to 0.8\,kpc by the violent merger event, while the latter population remains confined to a region of $\sim 0.1$\,kpc  where the gravitational potential of the post-merger halo is deepest. We discuss the origin of the metallicity gradient further below.

\section{Discussion}

We have run a set of cosmological galaxy formation simulations to show that a major merger among some of the earliest galaxies to have formed can generate a spatially extended stellar structure in and around a UFD. Specifically, an elongated stellar structure is formed in the direction of the merger. The bar-like structure extends over $\sim 1\ \mathrm{kpc}$, and it survives at least for $2.5\,\mathrm{Gyr}$ after the merger. The surface stellar density profile $\Sigma(R)$ is significantly altered by the merger. About $7$\%  of stars are then found at large radii, beyond $> 5\ r_\mathrm{half}$. In contrast, 
our simulated UFDs without mergers are compact, with almost no stars at $> 5\ r_\mathrm{half}$, which is consistent with an exponential surface stellar density profile. 

\subsection{Surface density profile}

The surface brightness of massive elliptical galaxies follows an empirical relation \citep{1948DeVaucouleurs}:
\begin{equation}
    I(R) \propto \exp\biggl[-7.669\biggl(\frac{R}{R_e}\biggr)^{1/4}\biggr].
    \label{eq:de vaucouleurs}
\end{equation}
Dwarf spheroidal galaxies tend to have more concentrated profiles with S\'ersic indices of $\sim 1$ \citep{2018Munoz_dSph_SurfaceDensityProfile} as also found for our galaxy samples that do not experience major mergers.
Galaxy-galaxy merger can significantly heat the system and change the density profile from a compact King profile \citep{1966King_profile} to an extended de Vaucouleurs profile (Eq.~\ref{eq:de vaucouleurs}) up to $\sim 10\ r_\mathrm{{half}}$ \citep{1986Aguilar_density_profile}. The resulting de Vaucouleurs profile is robust against mergers: a collision of two galaxies with de Vaucouleurs profiles produces the
stellar distribution consistent with a de Vaucouleurs profile. 
Our results show that the earliest galaxies with much lower masses follow the same trend. Mergers can dynamically heat the system and shape their surface density profile that resembles a de Vaucouleurs profile. We expect that the galaxies that do not experience mergers will remain compact even today, and that they constitute a population of  dwarf spheroidal galaxies with stellar components following exponential profiles.

\subsection{Origin of the metallicity gradient}

\citet{2021Chiti_TucIIhalo} report a metallicity gradient of $-0.87\pm -0.30$ dex kpc$^{-1}$ for Tuc~II. This broadly resembles the gradients found across all post-merger snapshots, as our merged galaxy consistently shows a metallicity gradient, although most gradients are slightly smaller than what is shown in Figure~\ref{fig:FeH_distance}. \citet{1980White_merger} discuss the emergence of metallicity gradient of galaxy merger remnants. In general, population mixing during the merger reduces the metallicity gradient compared to that before the merger. However, there is also a possibility that star formation after the merger can produce steeper gradient by forming metal-rich stars at the center.

The metallicity gradient in the merged galaxy appears to have formed in a series of steps. First, there is a small  difference of the average metallicity of 0.3\,dex between the stellar components of the central and infalling galaxies. The more metal-rich central galaxy remains largely undisrupted and its stellar component stays within the central region during the merger. In contrast, the more metal-poor, infalling galaxy is disrupted during the merger and its member stars are re-distributed to large radii within the merger remnant. By tracking the formation histories of these star particles, we find that most stars on the eventual distant orbits are actually formed during the merger process (see the bottom panel in Figure~\ref{fig:FeH_distance}). 
Because the infalling galaxy is relatively more gas-rich
than the central galaxy that has almost completed its star formation at the time of the merger, the fast encounter with the central galaxy generates strong shocks that lead to an episode of vigorous star formation.
Stars formed during the merger are metal-poor and have large kinetic (orbital) energies. 
In addition, stars are formed after the remaining gas has settled in the center region of the newly formed merger galaxy $\sim$ 300\,Myr after the merger. 
The stars tend to be more metal-rich owing to the additional chemical enrichment by massive stars formed earlier during the merger,
and they remain in the central region. The series of events described above finally shape the metallicity gradient as shown in Figure~\ref{fig:FeH_distance}.

The emergence of the metallicity gradient in the merged galaxy is overall similar to, but slightly different from the original picture of \citet{1980White_merger}. A steep metallicity gradient is produced in our simulation, in which
population mixing is incomplete between the two merging galaxies, 
but is efficient between the inner and outer stars within each galaxy. 

There may be an interesting connection between our model of the metallicity gradient formation and observations of classical dwarf spheroidals (dSphs) in Local Group in terms of the multiple stellar subcomponents (see \citet{2004_Tolstoy_Sculptor_multiple, 2011_Walker_Penarrubia} and references therein). Some dSphs show multiple stellar components with distinct metallicities within each system. For these dSphs, the metal-rich component is concentrated in the center while the metal-poor component shows lower concentration. \citet{2016Benitez_Llambay} study the formation of such multiple stellar subcomponents in classical dSphs using cosmological simulations. 
The unusually extended stellar profile and the tentative evidence of metallicity gradient in Tuc~II may provide a hints for the existence of multiple stellar populations. Further observation that clarifies the existence of distinct stellar in UFDs components would be crucial to give a clear answer.

\subsection{Long-time evolution of an early merged galaxy}

We stop the simulation at 2.5\,Gyr and before the merged galaxy would get accreted by a larger halo. When we make a comparison with present-day UFDs such as Tuc~II, 
a number of processes need to be considered that could affect the galaxy's
structural properties over time. 

In the following, we consider the possibility that the merged galaxy would fall into a Milky Way-mass halo at some later epoch. We  discuss the potential effects of (i) dynamical friction, (ii) relaxation, and (iii) tidal interaction with the host halo or with other satellite galaxies that our simulation UFD might eventually experience. 

The dynamical friction time $t_\mathrm{fric}$ for a UFD with mass $M$ that resides in a host halo whose velocity dispersion is $\sigma$ is
\begin{equation}
    t_\mathrm{fric} = \frac{19\,\mathrm{Gyr}}{\mathrm{ln}\,\Lambda}\biggl(\frac{r_{i}}{5\,\mathrm{kpc}}\biggr)^{2}\frac{\sigma}{200\,\mathrm{km\,s^{-1}}}\frac{10^{8}\,\Msun}{M},
\end{equation}
where $r_i$ is the initial distance from the center, and $\mathrm{ln}\,\Lambda \simeq 6$ is the Coulomb logarithm \citep{2008_GalacticDynamics}. Equating this with the age of the Universe of 14 billion years, we obtain the initial radius $r_{i}$ of $\sim 10\,\mathrm{kpc}$ as the critical radius for a UFD to sink to the center
by the dynamical friction. In general, the dynamical friction on a UFD would be significant only after it comes very close to the center of the host halo.


As a simple estimate for the effect of a non-smooth gravitational field generated by UFD member stars, we calculate the two-body relaxation timescale. 
The relaxation time for a UFD with $N = 10^{4}, m = 1 \Msun, M=10^{8} \Msun, R = 0.1 R_{vir}, \log \Lambda = 10$ is given by
$t_\mathrm{relax} = (8N(Gm/R)^{2}v^{-4}\log \Lambda)^{-1}t_\mathrm{cross}$,
which yields $3\times 10^{15}$\,yr.
These estimates suggest that neither dynamical friction nor relaxation impacts the structure of a UFD surviving for an extended period of time.

The third process, tidal interaction with the host halo, may increase the half-light radius by further heating the system in case of cored dark matter halos. After infall, a UFD might be heated or disrupted by the tidal field of the host halo. Indeed, some UFDs that orbit close to the galactic center show signatures of tidal disruption, e.g., Tucana~III with an estimated pericentric distance of 3\,kpc (\citealt{2017Simon_TucIIIobs, 2018Li_TucIII_stream}). However,  Tuc~II has an estimated pericentric distance of 39\,kpc, although this value depends on the mass estimate of MW halo. For example, \citet{2018Fritz} estimate the pericenter to be 29\,kpc assuming a mass of MW halo of $1.6\times 10^{12}$ \Msun. In any case, the orbit of Tuc~II does not approach very close to the center of Milky Way (\citealt{2018Simon_Gaia_UFD}).  
\citet{2008Penarrubia} show that the stellar component is more resilient to tides than the dark matter component, owing to the considerably smaller size and mass. Considering the presence of the dark matter component in Tuc~II today, we expect that tidal heating has not been significant in shaping its overall structure. It is, however, possible that the stellar envelope of a UFD created by a past merger is more susceptible to tidal forces than in the case of the centrally-concentrated stellar profile. We argue that, particularly for Tuc~II, such stripping was unlikely to be important, considering the fact that the farthest observed star in the outskirt of Tuc~II is gravitationally bound.


\subsection{Chemical enrichment}

An interesting feature of Tuc~II is that it has at least one chemical abundance outlier, Tuc~II-033 \citep{2018Chiti_TucII}, among the stars studied in detail with high-resolution spectroscopy. This star has distinctly high [Sr/H] and [Ba/H] abundances, paired with a low [$\alpha$/Fe] value compared with other member stars. This could indicate a delayed contribution by AGB stars and by type-Ia SNe in Tuc~II. 

Our simulation has shown that a galaxy merger plays an important role in explaining the observed, spatially extended stellar population of the UFD Tuc~II. 
Although star formation in a UFD is typically limited due to its short formation history, galaxy mergers naturally introduce another or prolonged episode of star formation. 
As a consequence, stars formed during such a merger have higher [Fe/H] abundances and also higher amounts of {\it s}-process elements \citep{2020Tarumi_UFD_sprocess}. $S$-process enrichment by AGB stars that typically arise on longer timescales can be traced with Ba abundances. Applying this to Tuc~II-033 suggests that its Ba abundance could be reproduced if AGB yields were $\sim 0.5$\,dex higher than what was assumed in our more extended chemical evolution study of this early galaxies \citep{2020Tarumi_UFD_sprocess}. There we have already found that there is an overall higher Ba content among stars formed in the merger compared to pre-merger stars that display lower Ba abundances. 

It appears that Tuc~II's longer term chemical evolution has been influenced by its early merger. However, there are other possibilities for high observed neutron-capture abundances, and the high Sr abundance in Tuc~II-033 cannot be easily explained with AGB yield alone. An additional contribution by the so-called light-element primary processes, e.g., an electron-capture supernovae \citep{2018Wanajo_ECSNe}, may have operated independently of low and intermediate mass AGB stars that contributed light neutron-capture processes such as Sr.

In conclusion, assuming that Tuc~II-033 is not a halo interloper star contaminating the Tuc~II sample, Tuc~II appears to be an example of an ancient wet-merger remnant of two primitive galaxies. Additional measurements of neutron-capture element abundances of all the known Tuc~II stars, especially those at large radii, would be helpful to further map the possible late-time chemical evolution of this galaxy.

\subsection{Galactic building blocks} 

UFD galaxies have long been speculated to be candidates for surviving 
early galactic building blocks given their low masses and metallicities, and old ages \citep{2012Frebel_FirstGalaxies, Simon19_UFDreview}.
The extended halo of Tuc~II and the fact that our simulation of an early merger of 
even more primitive, low metallicity galaxies can principally explain Tuc~II's overall structure 
calls into question that Tuc~II, and perhaps other UFDs, are surviving building blocks. 
Instead, our simulation suggests that the detailed structure of an UFD provides 
information on the past events and on the type of progenitors that have shaped the UFD.
Clearly, future searches for extended structures around other UFDs are needed.

Our final merged galaxy has a total mass of $2.5\times10^8$\Msun, suggesting that Tuc~II may have been slightly more massive than today as its mass is now estimated to be at least $10^7$\,\Msun\,at 1 kpc from the center \citep{2021Chiti_TucIIhalo}.
It thus appears suggestive that early stellar systems with then-total masses of 5 to $9\times10^7$ \,\Msun\, and then-tiny stellar components of 4,000 to  8,000\,\Msun\, are plausible candidates of 
building blocks of subsequent galaxy formation.

\subsection{On the fraction of early merged galaxies with extended stellar profiles}
We show that a major merger of two primordial galaxies results in a system with an extended stellar profile. We have run additional simulations to estimate the occurrence rate of such early mergers. Among a total of 15 galaxies we have simulated, we find that only one experiences a major merger with the result of developing an extended stellar surface density profile. 
Hence our rough estimate of the fraction is $\sim 10$ percent, although this should be taken as a lower limit. The fraction would be higher if we consider other environments such galaxies in over-density regions.
Also, since we stop the simulation at $z=4$, possible later time mergers with either field galaxies or with satellite galaxies are not followed. Some of our simulated galaxies might
develop an extended stellar structure through the late-time events.

To address the important question more quantitatively, simulations with a larger volume would be needed that contains a large number of small building block-type galaxies,
while resolving star formation in individual galaxies. The fraction of early merger systems remains to be an important quantity to compare with observations that will assess the number of surviving UFDs with extended stellar profiles such as Tuc~II. 


\section{Summary and Conclusion}  

We have performed cosmological simulations of the formation of the earliest galaxies in the universe. Specifically, we have investigated a merger of two building block-type first galaxies. The early galaxy merger dynamically heats the emerging system, induces star formation and 
produces a spatially extended stellar halo. 
The radial density profile of the merged galaxy roughly follows a de Vaucouleur profile 
whereas other, non-perturbed systems have exponential profiles. 

Our simulation reproduces the observed spatial distribution and [Fe/H] abundance gradient of the ultra-faint dwarf galaxy Tucana~II that was found to have an extended stellar halo with stars being away from the center up to 1\,kpc, and an overall metallicity of $\mbox{[Fe/H]}\sim-2.8$ \citep{2021Chiti_TucIIhalo}, with the more distant stars being more metal-poor.

We propose that future observations of the vicinity of known UFDs in search of similarly extended halos will provide valuable information on the formation and evolution of these systems in the early universe. If observed stellar density distributions are well-fitted with a centrally concentrated, exponential profile, the galaxy likely did not experience any significant merger in the past, and are likely (candidates for) surviving first galaxies \citep{2012Frebel_FirstGalaxies}. De Vaucouleur profiles would point to a violent assembly history instead, as may be the case for Tuc~II.

Further theoretical studies on effects such as tidal heating and on early galaxy mergers with a variety of properties would help to understand the formation mechanism of spatially extended structure. It is also important to explore how large variation is expected among known UFDs. This will help to better understand the nature and formation processes of the early galaxies, which mark the beginning of hierarchical assembly. More generally, such studies will reveal how large galaxies like Milky Way grew from these galactic building blocks.

\acknowledgments

We thank the anonymous referee for providing constructive comments.
Numerical computations were carried out on Cray XC50 at Center for Computational Astrophysics, National Astronomical Observatory of Japan.
Y. T. is supported by JSPS KAKENHI Grant Number 20J21795.
A.F. acknowledges support from NSF grant AST-1716251, and thanks the Wissenschaftskolleg zu Berlin for their wonderful Fellow's program and generous hospitality.


%

\vspace{5mm}


\software{matplotlib \citep{Matplotlib}, numpy \citep{Numpy}, scipy \citep{SciPy}}




\bibliography{rprocess}{}
\bibliographystyle{aasjournal}


\listofchanges

\end{document}